\documentclass[nologo,11pt,a4paper]{ETHpaper} 
\usepackage{graphicx,subfigure}
\usepackage{hyphenat}   
\usepackage{url} 
\usepackage{amsmath} 
          
\begin{document}  

\title{A Network Perspective on Software Modularity}

\titlealternative{A Network Perspective on Software Modularity}

\author{Marcelo Serrano Zanetti and Frank Schweitzer} 

\authoralternative{Marcelo Serrano Zanetti and Frank Schweitzer}

\address{Chair of Systems Design, ETH Zurich, Switzerland\\ \url{http://www.sg.ethz.ch}}
 
\reference{ARCS Workshops 2012, pp. 175--186}

\makeframing

\maketitle

\begin{abstract} 
Modularity is a desirable characteristic for software systems. In this article we propose to use a quantitative method from complex network sciences to estimate the coherence between the modularity of the dependency network of large open source \textsc{Java} projects and their decomposition in terms of \textsc{Java} packages. The results presented in this article indicate that our methodology offers a promising and reasonable quantitative approach with potential impact on software engineering processes. 
  
\end{abstract}   
   
\section{Introduction}   

\label{intro}
  
     
 The modularity of a software architecture is considered a key feature that contributes to the sustainability of large scale software projects \cite{modularity1985}. Ideally, modularization fosters the decoupling of software development efforts, which can then be performed independently if a binding standard interface is established. As the software evolves in time, modularity might even favor its maintainability and expandability. If the development of a given system is meant to be sustainable, the amount of effort required to perform modifications in the software architecture must be compatible with the resources (time, human, etc) available at any time. Therefore monitoring the modularity of an evolving software system promises to be an important step towards a sustainable software development regime, however such a task would be tedious and slow if performed manually.
 

 In this article we propose an efficient automatic quantitative approach to estimate the coherence between the modularity of the dependency network of large open source \textsc{Java} projects and their decomposition in terms of \textsc{Java} packages. Our method is based on the well-established complex networks framework \cite{barabasinetsreview2002}\cite{newmannetsreview2003}. In order to adopt this framework, the first necessary step is to restate software modules and software systems in terms of network structures (see \cite{icsesoftwaredependencegraph1992}\cite{presoftwarenetworks2003}\cite{k09}\cite{gs11}\cite{msz11}).   
     
  
Through a network perspective, it is straightforward to visualize that the expected functionality of a software module is provided by the cooperation of fundamental software entities (functions, classes, procedures, etc) which perform the necessary operations. Thus a software module is a mesoscopic abstraction for a collection of entities acting microscopically. At the mesoscopic scale, software modules themselves become interdependent when integrated into a software system. Therefore the challenge in modularization of software consists in clustering highly dependent microscopic software entities, which are then packaged into software modules by minimizing the number of dependencies across modules after a system integration. This can be directly mapped to the software engineering literature, where modularity is defined as a high degree of intra-module \emph{cohesion} and low inter-module \emph{coupling} \cite{se2003}. As an example, since the number of dependencies across modules is expected to be minimized, a modular system is relatively easy to be upgraded through the replacement of an obsolete software module by a new one.   
   
          
Our contribution is based on a quantitative metric that measures the coherence between the decomposition of a software system into software modules and the cluster structures found in the network model of the software at a microscopic scale. However, 
here we do not attempt to construct module mappings that optimize this coherence. 
We only monitor the modularity of a software system already decomposed in terms of software modules. For this, we use a quantitative metric which describes a macroscopic property of a system composed of microscopic and mesoscopic structures (software dependencies and modular decomposition respectively). In other words, our method can measure the global impact of modifications made locally during the time evolution of a given software project. To illustrate the dynamics of this process, we study the time evolution of the degree of modularity expressed through our method for 28 open source \textsc{Java} projects. Our dataset contains different versions of the source code which were extracted periodically from the respective online software repositories. We argue that the application of the complex systems framework in the study of software systems provides valuable insights into the software engineering processes and the sustainability of large scale software projects.  
  

In section \ref{met} we present the details of our implementation and approach. Section \ref{res} discusses our preliminary results and in section \ref{rel} we comment on related work. Finally, in section \ref{conc} we conclude our work and we then elaborate on further research ideas. 
  
\section{Methodology}
\label{met} 
 
 
The starting point of our methodology is the re-expression of source code dependencies in terms of network structures. Conceptually, such an approach will differ for the targeted programming language and programming paradigm. 

We choose to focus our efforts on software written in \textsc{Java}, for it is an object-oriented programming language which suggests a straightforward re\hyp interpretation in terms of networks: \textsc{Java} classes are taken as network nodes, while a network edge will connect any two nodes if the corresponding \textsc{Java} classes share at least one software dependency (call, access of property, inheritance, etc). Another relevant aspect of \textsc{Java} is its built-in support for software modularization through the assignment of classes to packages. Last but not least, \textsc{Java} is a very popular programming language among free and open source software developers, and therefore plenty of examples containing the complete source code evolution is available online in software repositories and web software development platforms, such as \textsc{GitHub}\footnote{\url{https://github.com/}} and \textsc{sourceforge}\footnote{\url{http://sourceforge.net/}}. 
  
\begin{figure}[htpb] 
\begin{center}
\includegraphics[width=\textwidth]{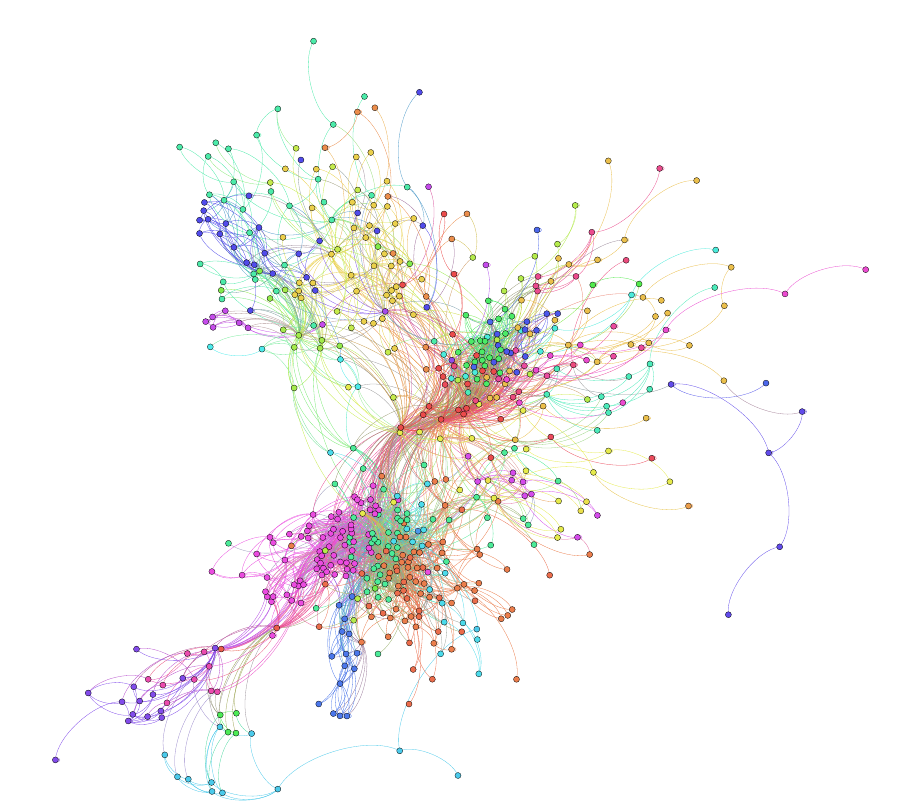}      
\caption{Visualization of the modular network structure of \textsc{AspectJ} as of 01-Aug-2004 (only the largest set of nodes connected via direct or indirect edges - largest connected component). This visualization was generated by \textsc{Gephi} \cite{gephi}.}
\label{f:aspectjlcc} 
\end{center}     
\end{figure}

Figure \ref{f:aspectjlcc} presents a visual example of the software network resulting from the application of the aforementioned method to one of the versions of the source code of \textsc{AspectJ}, which is a \textsc{Java} framework supporting the implementation of software using the aspect-oriented programming paradigm. In our dataset, this network grows from 654 up to 1651 nodes (classes). In this example, each color represents the package membership (module) of each class found in source code. This network perspective on source code can be extended in a relatively easy way to other programming languages and paradigms. See \cite{icsesoftwaredependencegraph1992}\cite{presoftwarenetworks2003} for more examples and approaches. 
    
As demonstrated in Figure \ref{f:aspectjlcc}, the visualization of network structures is a very useful technique for the analysis of the modularity of a given software architecture. However, a quantitative approach is still desirable since it allows us to capture the structural organization of a network in terms of a single numeric measure. This can be used to analyze the time evolution of a modular software architecture and can also be applied in a statistical correlation analysis when considering different quantitative metrics.
      
\begin{figure}[htpb]  
\begin{center}
\includegraphics[width=0.23\textwidth]{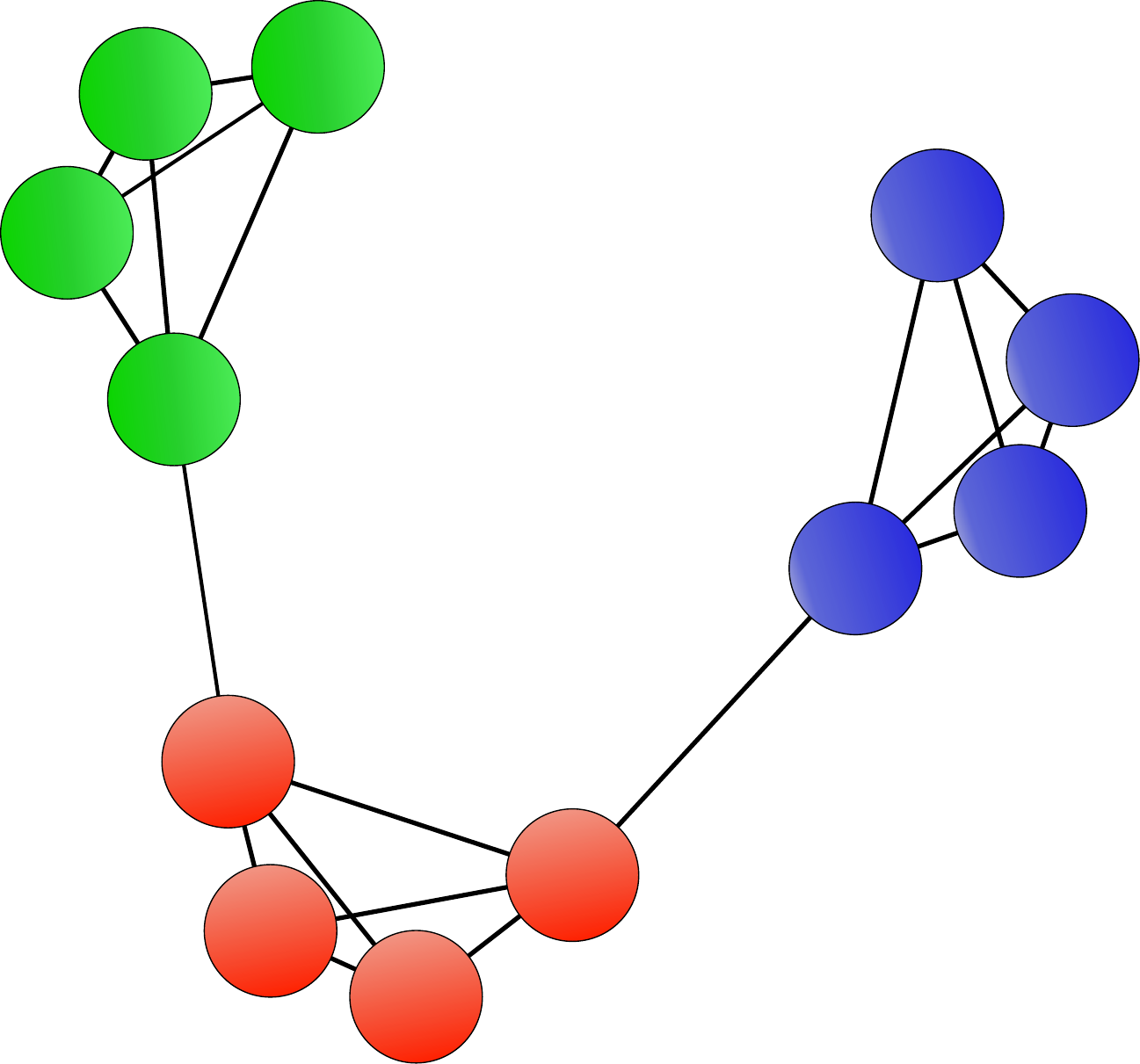} \includegraphics[width=0.23\textwidth]{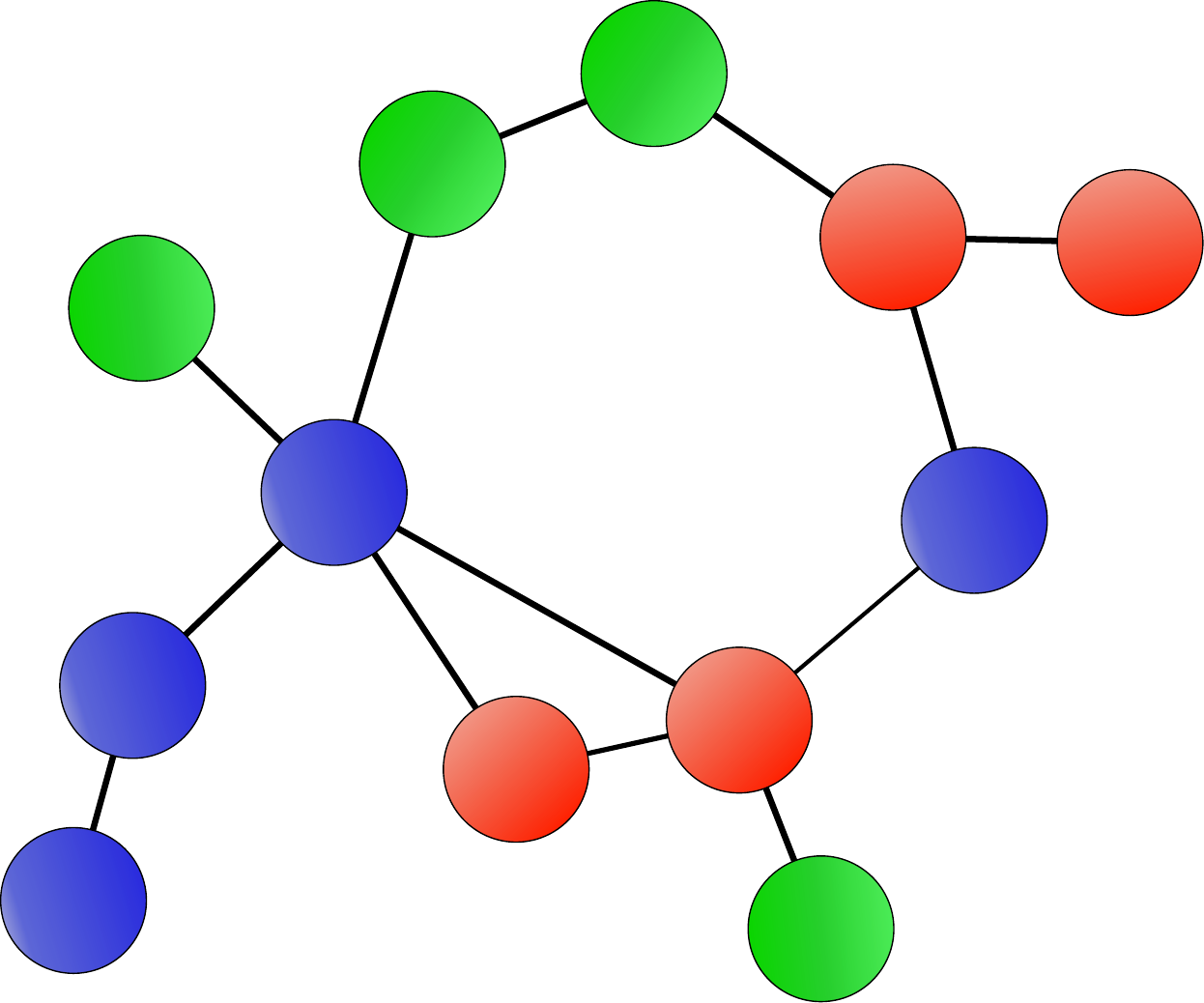}
\caption{Two examples of undirected networks where nodes (circles) with the same color are part of the same module. (left) modular network Q=0.8499. (right) random connectivity Q=0.0545.}
\label{f:exs} 
\end{center}               
\end{figure}

In recent years, the network sciences community has developed a number of quantitative metrics which capture structural features like e.g. clusters as well as the impact of nodes, clusters or any other structural entities on dynamical processes like e.g. information or failure spreading, consensus, opinion formation or synchronization \cite{newmannetsbook2010}. According to our needs, we adopt a network metric which was first used to study assortative mixing in networks, which is the tendency for network nodes to be connected to other
nodes that are like (or unlike) them in some way \cite{newmanR2003}. Assuming that sharing the same module membership makes nodes alike (and unlike otherwise), this metric could then be used to measure the modularity of network structures \cite{newmanQ2004}. 
For a given definition of \emph{modules} or \emph{clusters} and their underlying network structure, its respective degree of modularity is defined by

\begin{equation}    
\centering
Q=\frac{\sum_i^n{e_{ii}}-\sum_i^n{a_ib_i}}{1-\sum_i^n{a_ib_i}}
\label{eq:q}
\end{equation}   

where $e_{ij}$ is the fraction of all edges in the network that link nodes in module $i$ to nodes in module $j$, $a_i=\sum_j^ne_{ij}$, $b_i=\sum_j^ne_{ji}$ (column and row sum respectively) while $n$ is the total number of existing modules. If the network is an undirected graph the matrix defined by $\textbf{e}$ is symmetric and $a_i=b_i$ \cite{newmanR2003}. The metric defined by equation (\ref{eq:q}) measures the fraction of network edges that connect nodes within the same module ($\sum_i^n e_{ii}$) minus the expected value of the same quantity measured from a random network with the same node/module allocation ($\sum_i^n a_ib_i$). If the first is not better than random $Q=0$ \cite{newmanQ2004}. However, $Q$ would not be defined if all edges are concentrated within a single module because the scaling factor $1-\sum_i^na_ib_i=0$ (no modular structure). In such a case we define $Q = 0$ as well. In general, $Q\in[-1,1]$, i.e. the more modular the network, the closer $Q$ is to $1$. Figure (\ref{f:exs}) provides two examples of networks and their respective $Q$ scores.       
         
In the analysis of software structures, this metric is useful because in many cases the definition of modules is given by means of programming constructs like classes, files, namespaces or packages. The $Q$--metric can thus be used to study how well the cluster structures in the network of dependencies correspond to the modular decomposition of a project in terms of packages, namespaces, etc. We applied the $Q$-metric in an analysis of the evolution of the modularity of the software architecture of a set of \textsc{Java} open source projects and we discuss our preliminary results in section \ref{res}.   
        
\section{Preliminary Results\label{res}}
  
Our analysis is based on a dataset containing the detailed time evolution for the source code of 28 open source \textsc{Java} projects. The snapshots of the source code of each project were extracted from the respective \textsc{cvs} online software repositories, on a monthly basis. Table \ref{t:prjs} displays the recorded period for each project. Most of those projects are hosted at \textsc{sourceforge} and were selected because they were the largest (number of classes) at the time the dataset was collected. The single exception is \textsc{Eclipse}, which has its own online facilities\footnote{\url{http://www.eclipse.org}}. The source code for \textsc{Eclipse} was thus obtained through a different setup. For each project the \textsc{cvs} change history and class dependence structure were extracted, processed and stored in a directed graph format, i.e. ($c_1$, $c_2$, $T$) which reads as \emph{$c_1$ depends on $c_2$ at time $T$}.

\begin{table}[htpb]
\begin{center}
\caption{The 28 \textsc{Java} projects which compose our source code evolution dataset. Most of those projects were extracted from the respective \textsc{cvs} software repositories hosted by \textsc{sourceforge}.}        
\label{t:prjs}  
{\small    
\begin{tabular}{lll | lll}
\hline\noalign{\smallskip}
project name & record start & record end & project name & record start & record end\\
\noalign{\smallskip}
\hline    
\noalign{\smallskip}
architecturware&2004-04-01&2007-12-01&jnode&2003-06-01&2005-12-01\\
aspectj&2003-01-01&2008-02-01&jpox&2003-09-01&2006-12-01\\
azureus&2003-08-01&2008-01-01&openqrm&2007-04-01&2008-03-01\\
cjos&2000-11-01&2007-12-01&openuss&2003-06-01&2006-12-01\\
composestar&2003-12-01&2005-12-01&openxava&2004-12-01&2007-12-01\\
eclipse&2001-05-01&2008-03-01&personalaccess&2004-11-01&2007-12-01\\
enterprise&2002-11-01&2007-12-01&phpeclipse&2002-08-01&2007-12-01\\
findbugs&2003-04-01&2007-12-01&rodin-b-sharp&2005-11-01&2007-12-01\\
fudaa&2003-02-01&2007-12-01&sapia&2002-12-01&2007-12-01\\
gpe4gtk&2005-08-01&2006-12-01&sblim&2001-07-01&2007-12-01\\
hibernate&2001-12-01&2005-12-01&springframework&2003-03-01&2007-12-01\\
jaffa&2003-03-01&2007-12-01&squirrel-sql&2001-12-01&2007-12-01\\
jena&2001-02-01&2008-02-01&xmsf&2004-02-01&2007-12-01\\
jmlspecs&2002-03-01&2007-12-01&yale&2002-04-01&2008-02-01\\
\hline  
\end{tabular}}
\end{center} 
\end{table} 
  
\begin{figure}[htpb] 
\begin{center}
\includegraphics[width=0.9\textwidth]{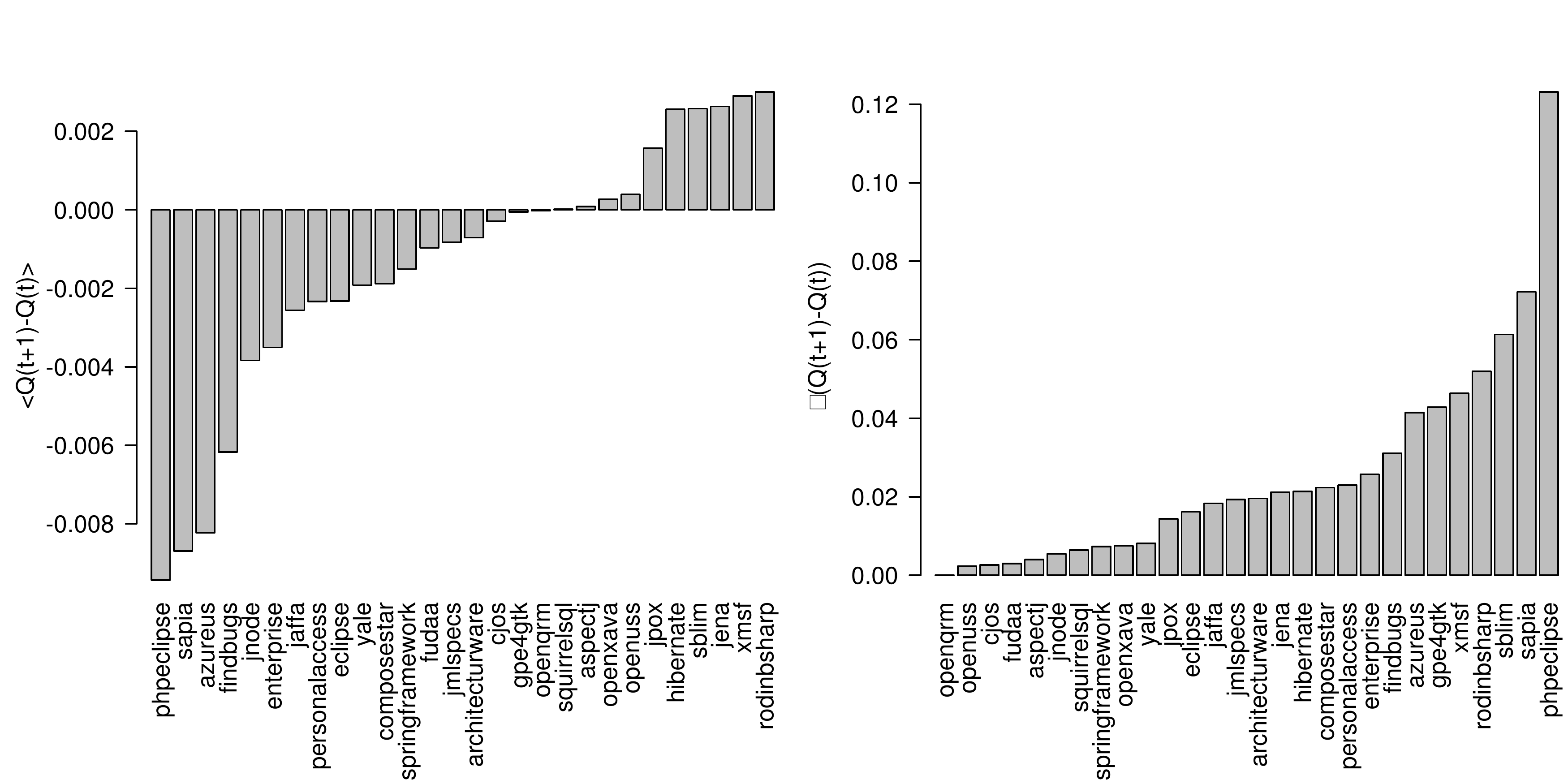}
\caption{Ranking software projects using the $Q$-metric. (left) ranking by average incremental change of the $Q$-metric over the observation period, estimated with <$Q(t+1)-Q(t)$>. (right) ranking by degree of fluctuation of the changes in the modularity over the studied period, estimated with $\sigma(Q(t+1)-Q(t))$.} 
\label{f:barplot}     
\end{center}
\end{figure}

Using the schema described in section \ref{met}, we applied the $Q$-metric to the network extracted from each snapshot within the recorded period. In order to facilitate the presentation of the time evolution of these projects, we first compose all projects into four groups, according to the degree of fluctuation of the $Q$--metric. In Figure \ref{f:barplot}, we thus compute the mean fluctuation in time of the $Q$-metric, i.e. $<Q(t+1)-Q(t)>$ where $t$ and $t+1$ are consecutive snapshots of the software and the average $<\cdot>$ is over all snapshots in the dataset. This approach captures the average incremental change of the $Q$-metric over the observation period. In the same figure, we also show the standard deviation of $Q(t+1)-Q(t)$, which captures the degree of fluctuation of the changes in modularity over the same period. We performed a ranking of projects along both the average incremental change and the fluctuations of modularity and these rankings are indicated in the abscissae of the respective plots (see Figure \ref{f:barplot}).  

 
\begin{figure}[htpb] 
\begin{center}   
\includegraphics[width=0.54\textwidth]{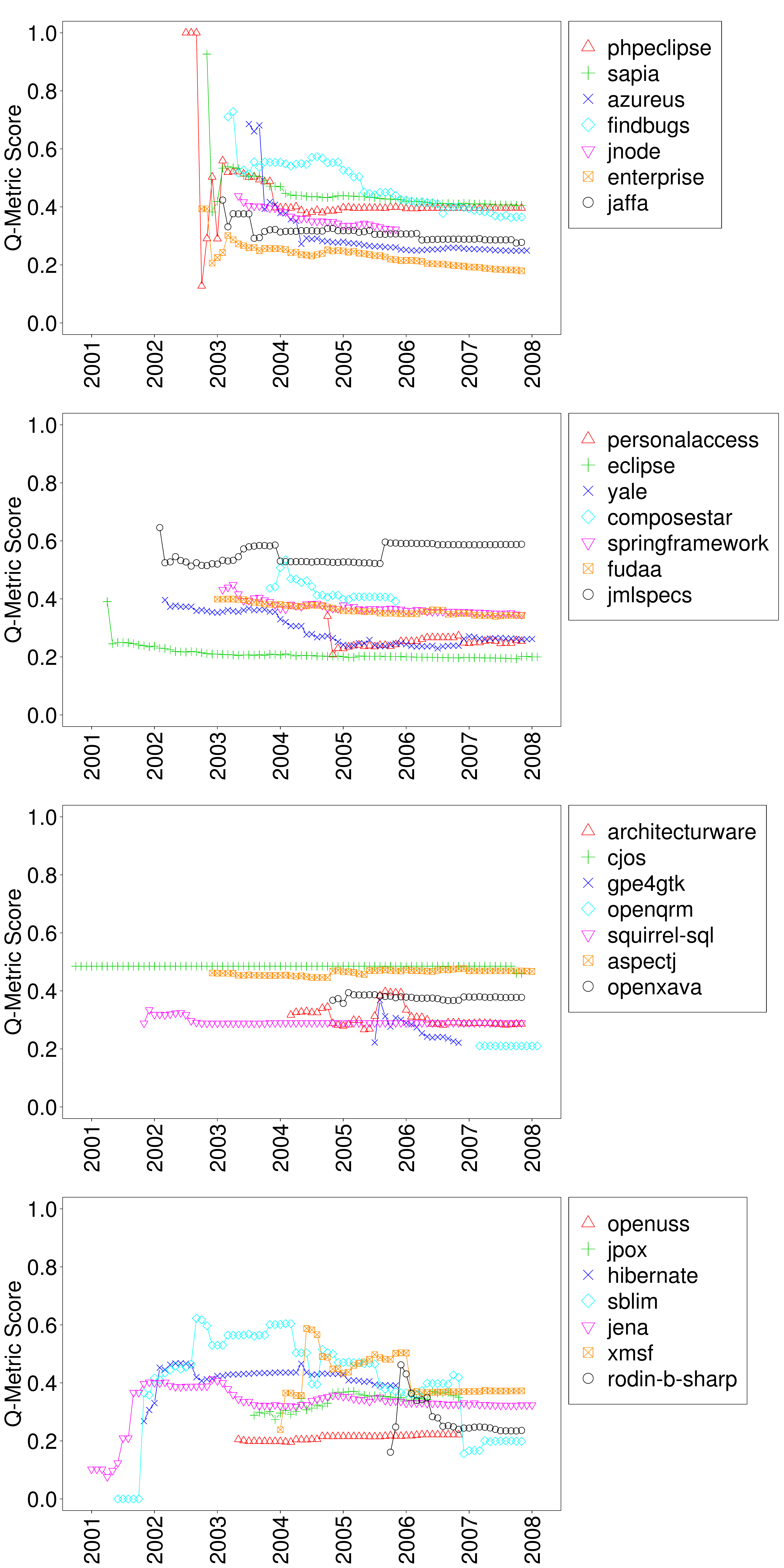}
\caption{Time evolution of the $Q$-metric score for each project in our dataset. The projects were sorted by the mean fluctuation in time of the $Q$-metric, i.e. $<Q(t+1)-Q(t)>$,  and displayed in increasing order of value (top-to-bottom). (top) highest mean decrease in $Q$. (bottom) highest mean increase in $Q$.}
\label{f:allqtimeav} 
\end{center}      
\end{figure}

The resulting plot, with the projects grouped and ranked by the average incremental change of $Q$ (see the left pannel of Figure \ref{f:barplot}), is shown in Figure \ref{f:allqtimeav}. Here, we observe that the $Q$-metric effectively classifies projects according to different dynamic regimes. In Figure \ref{f:barplot} (left) we can for instance focus on those projects that increase or decrease the software modularity, while Figure \ref{f:barplot} (right) can be used to study the most dynamical and the most stable software development regimes.
   
\begin{figure}[!htpb] 
  \centering
  \subfigure[\textsc{Azureus}]{\label{fig:EvolutionQ:Azureus}\includegraphics[width=0.3\textwidth]{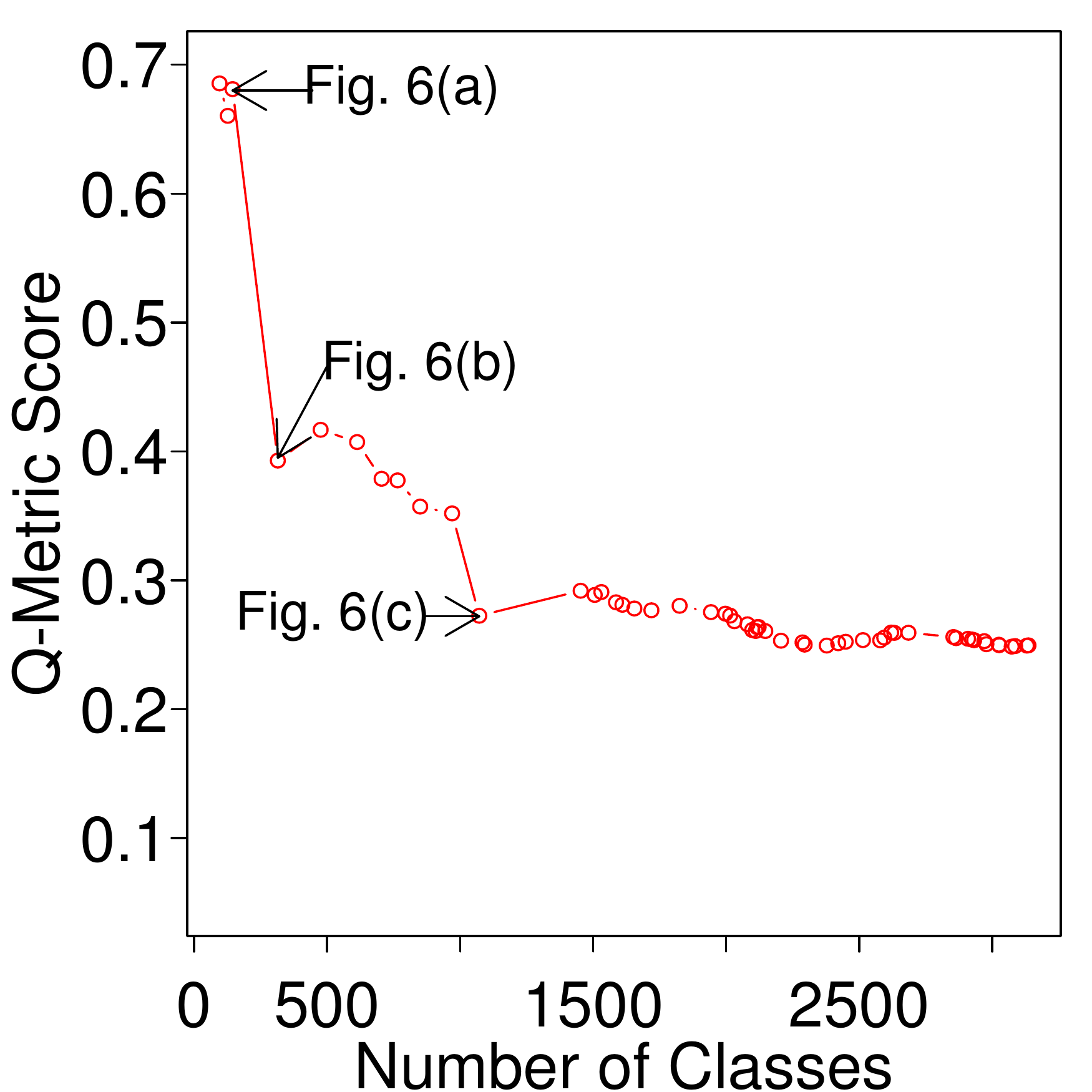}} \quad
  \subfigure[\textsc{Jena}]{\label{fig:EvolutionQ:Jena} \includegraphics[width=0.3\textwidth]{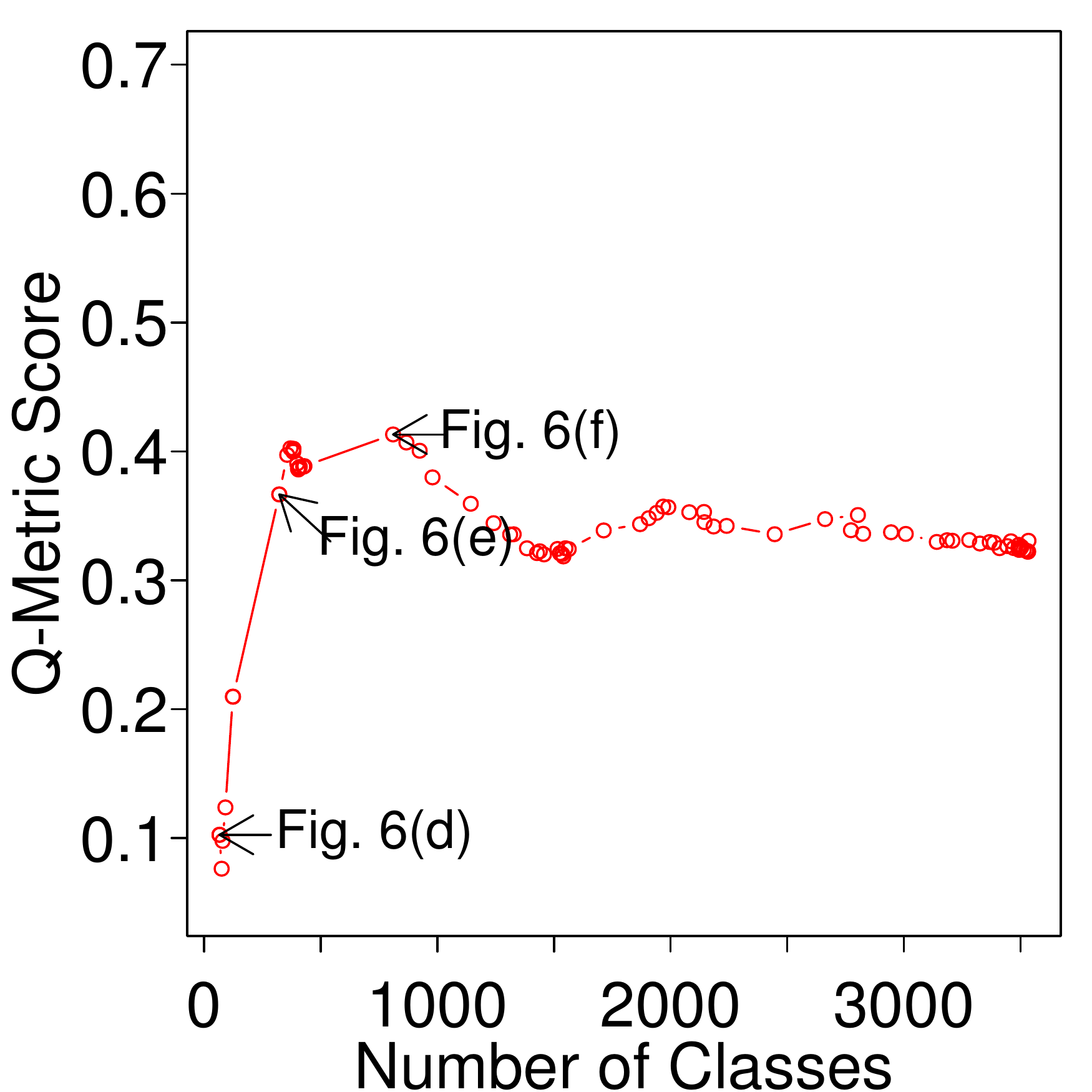}}
\caption{Detailed time evolution of the $Q$-metric for \textsc{Azureus} and \textsc{Jena}.}
\label{fig:EvolutionQ}    
\end{figure}

In the following we discuss two projects with contrasting evolution of modularity in more detail. In particular, for this we chose the projects \textsc{Azureus}, which is a torrent client being one of the projects with the largest average decrease in the $Q$-metric, as well as \textsc{Jena} which is a framework for building semantic web applications. In our dataset \textsc{Jena} actually shows one of the largest average increase of $Q$ (see the left plot in Figure \ref{f:barplot}). In Figure \ref{fig:EvolutionQ}, the time trajectory of the evolution of $Q$ is shown for both projects as a function of the total number of classes. As indicated in the Figures \ref{fig:EvolutionQ:Azureus} and \ref{fig:EvolutionQ:Jena}, three snapshots of the source code have been selected which cover the states of minimum and maximum modularity, as well as an intermediate state.
   
\begin{figure}[!ht]   
  \centering  
  \subfigure[2003-10-01]{\label{fig:azureus:01}\includegraphics[width=4cm]{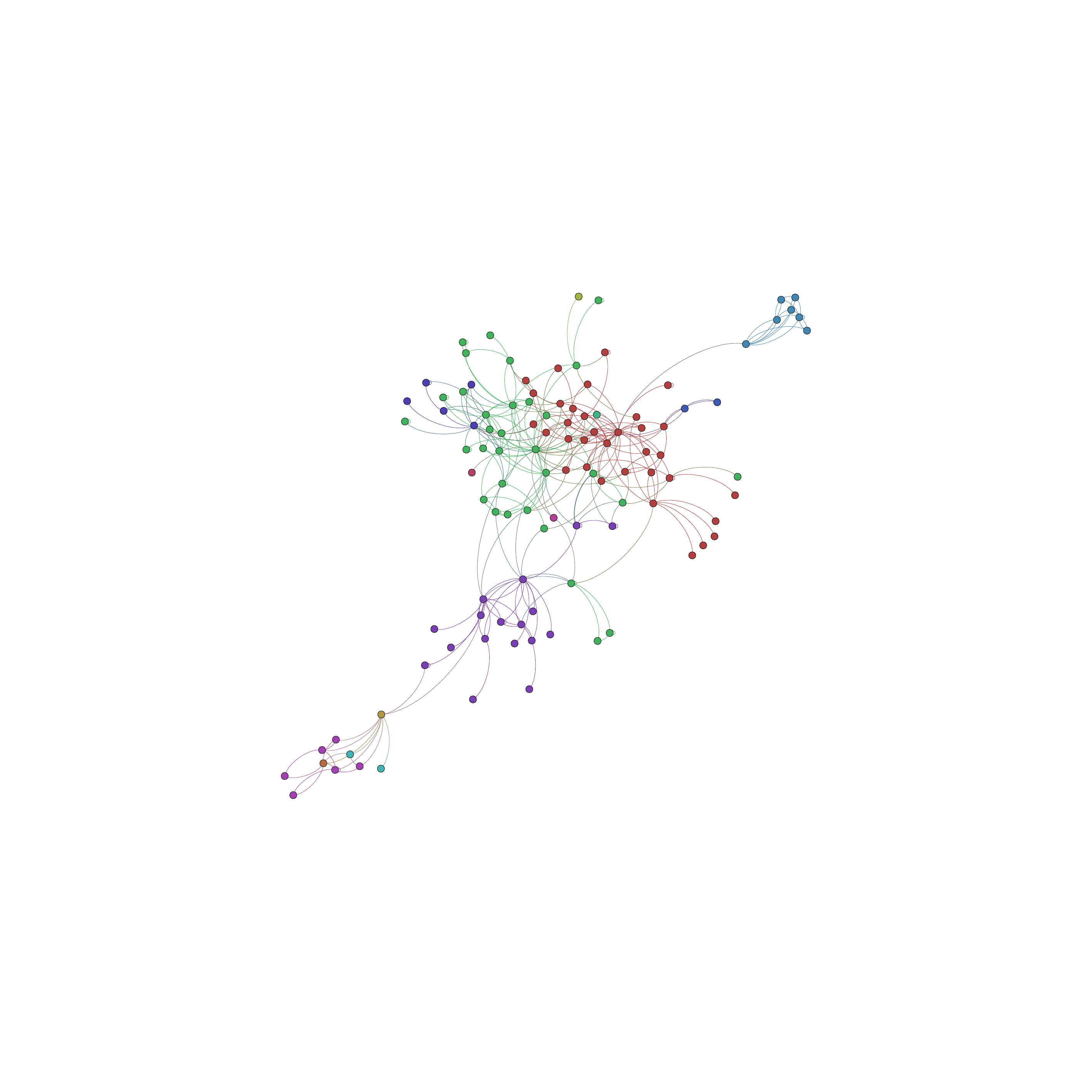}}
  \subfigure[2003-11-01]{\label{fig:azureus:02} \includegraphics[width=4cm]{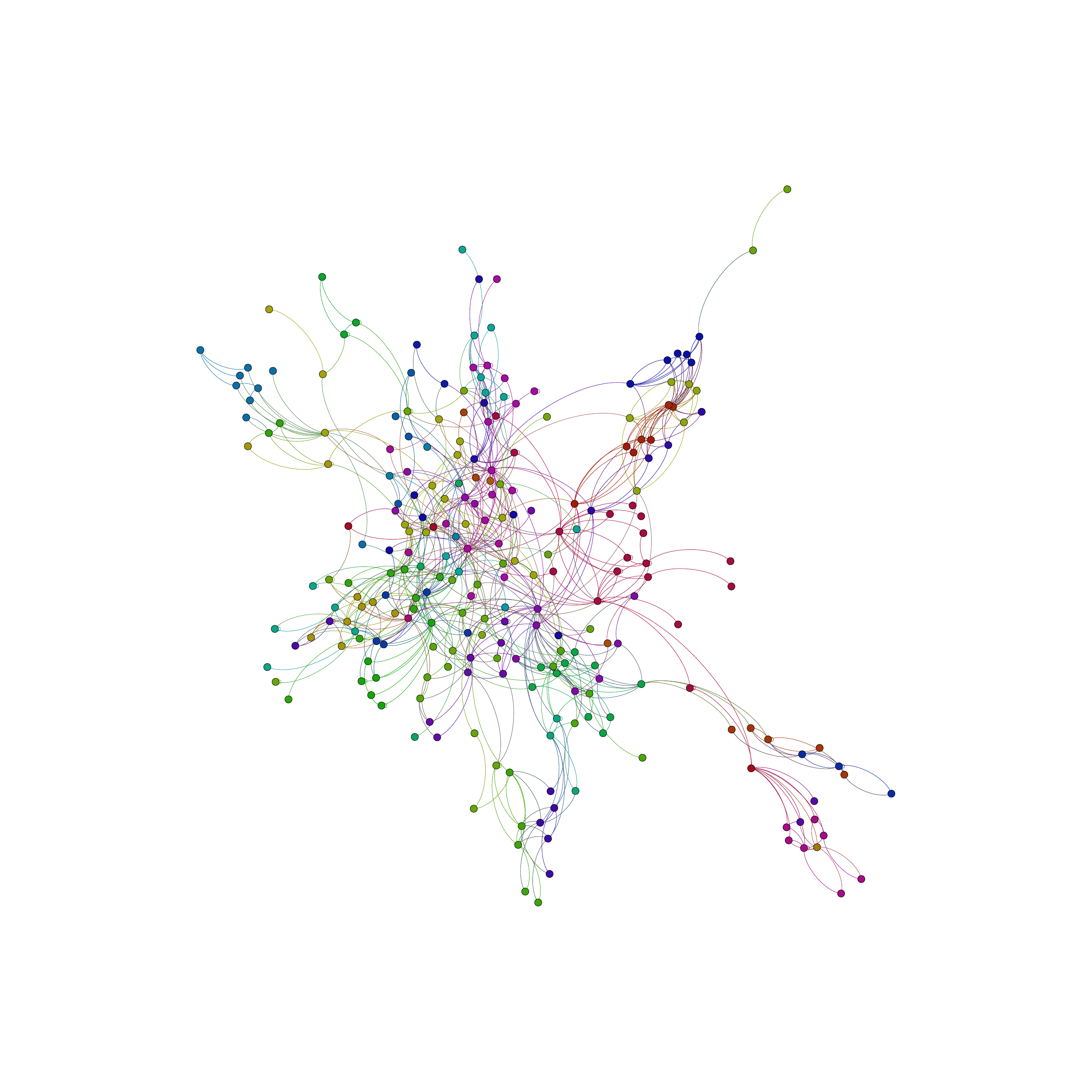}}
  \subfigure[2004-06-01]{\label{fig:azureus:03} \includegraphics[width=4cm]{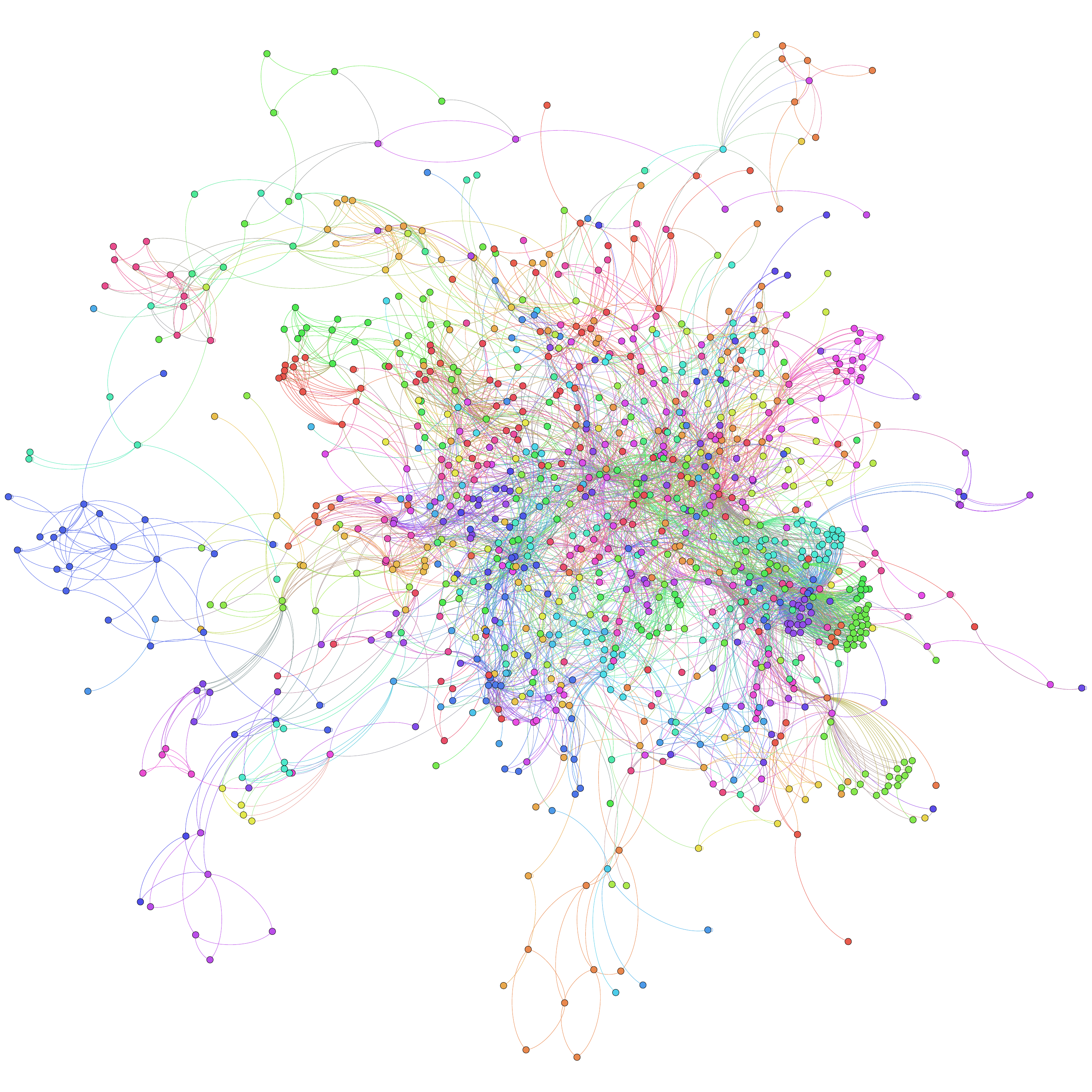}} \qquad 
  \subfigure[2001-02-01]{\label{fig:jena:01} \includegraphics[width=4cm]{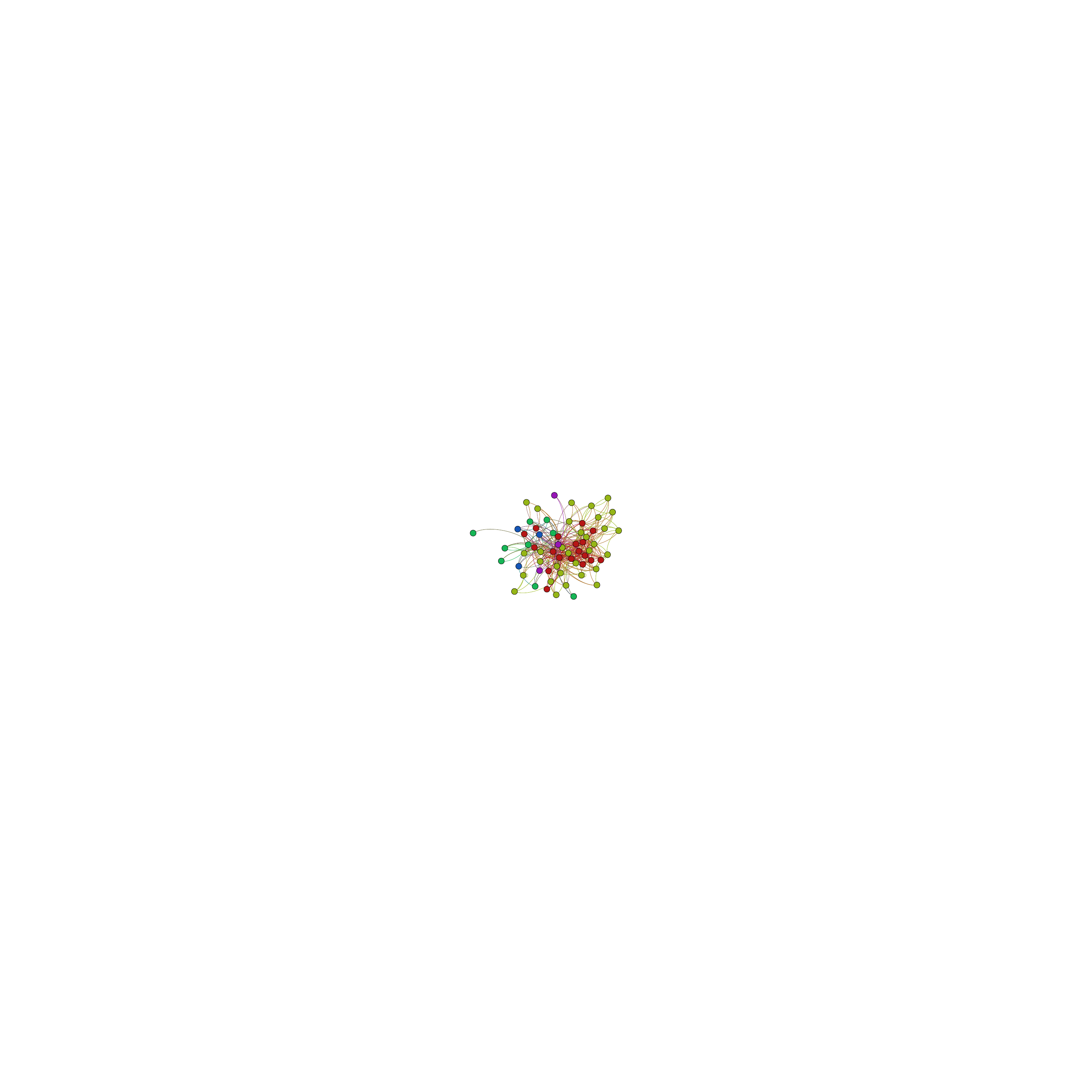}} 
  \subfigure[2001-10-01]{\label{fig:jena:02} \includegraphics[width=4cm]{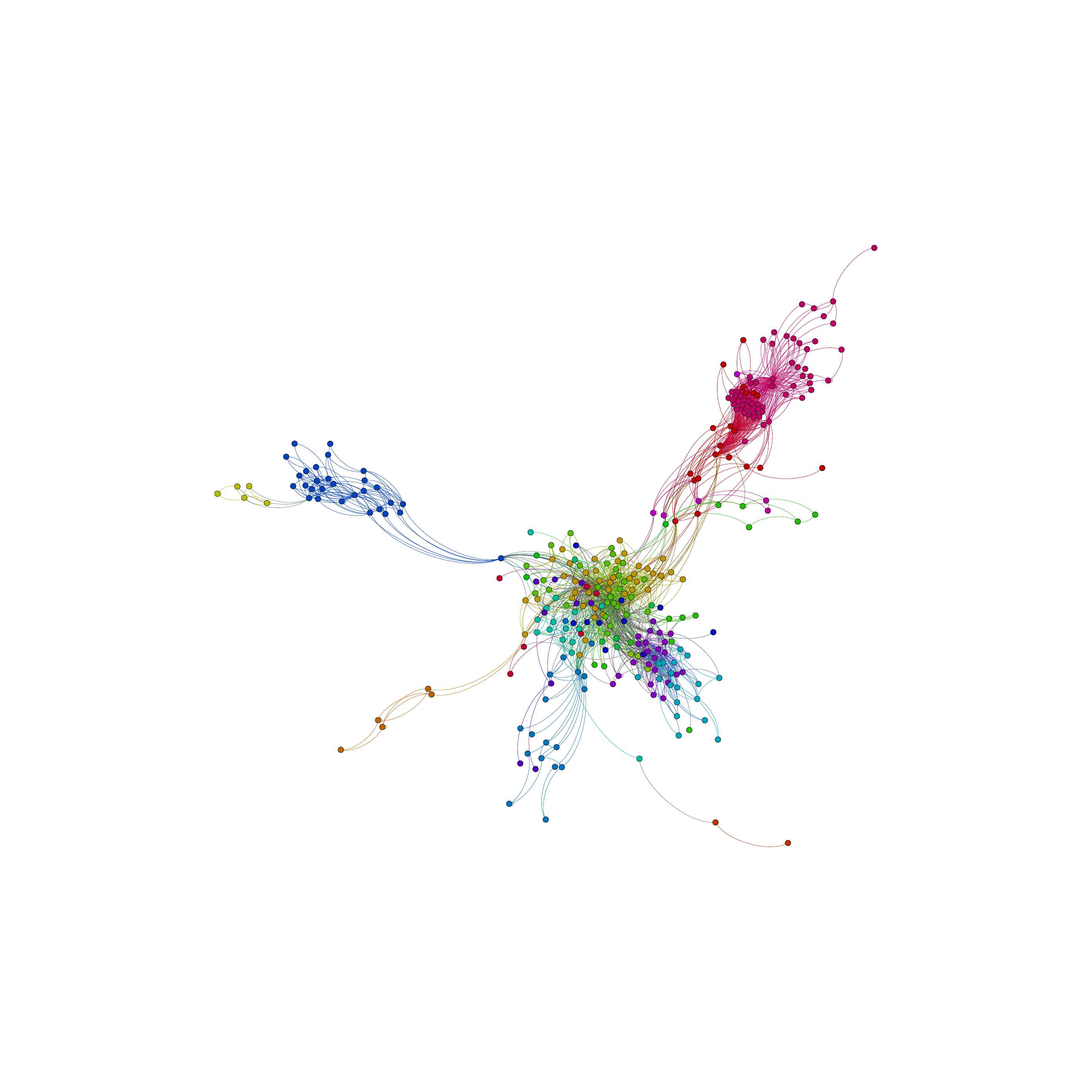}}
  \subfigure[2003-01-01]{\label{fig:jena:03} \includegraphics[width=4cm]{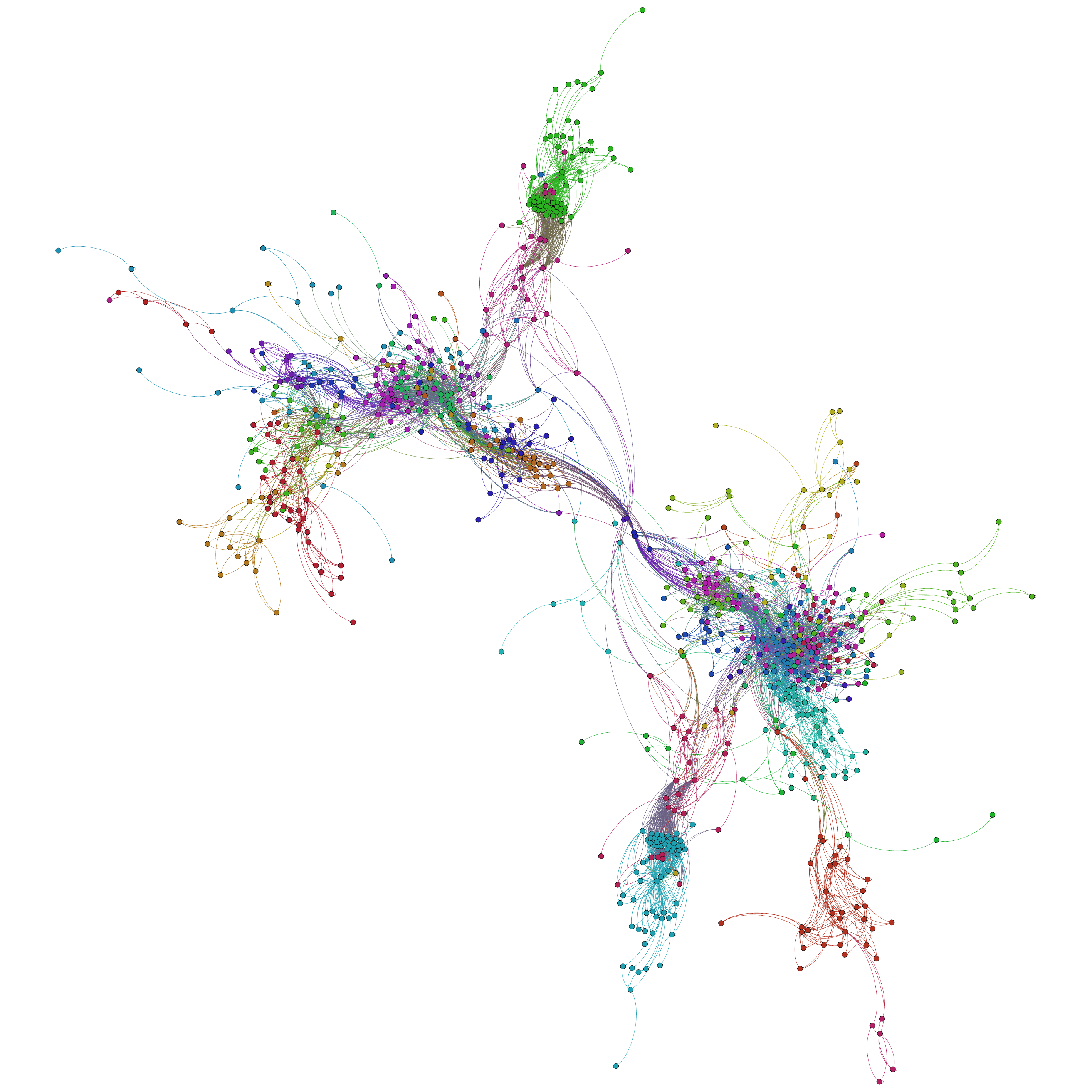}}
\caption{\label{fig:EvolutionNet}Three snapshots of the dependency networks of the projects \textsc{Azureus} (a-c) and \textsc{Jena} (d-f). Node colors in the individual networks indicate the decomposition in \textsc{Java} packages.}
\end{figure}
    
In Figure \ref{fig:EvolutionNet}, we show the dependency networks for the snapshots mentioned above. These networks have been created according to the methodology described in section \ref{met}, i.e. each node represents a \textsc{Java} class, while a dependency indicates a call, inheritance or usage relationship. Furthermore, nodes have been colored according to package membership. In order to visualize the coherence between the package decomposition of the classes and the modular organization of the dependency network, the networks have been layouted with the force-directed Yifan-Hu layout algorithm \cite{YifanHu2005}, which spatially organizes nodes according to cluster structures. In particular, nodes in networks with highly modular structures will be densely clustered in the resulting layouts and the modules will become clearly distinguishable. In the resulting networks we can visually examine how well the modular structures of the dependency network match the package structure of a project and thus obtain a visual impression of the module coherence expressed by the $Q$-metric. 
  
The effect of the different dynamical regimes in terms of the evolution of the $Q$-metric can easily be seen in the respective network structures. For the \textsc{Azureus} project, which is shown is Figures \ref{fig:azureus:01} - \ref{fig:azureus:03}, the coherence of the modular structure of the network of software dependencies with the package decomposition actually worsens over time, thus making it difficult to clearly separate packages in the resulting network structure. On the contrary, the evolution of the \textsc{Jena} project shows a very different dynamics. While the growth in terms of the number of nodes, packages and dependencies is in the same order of magnitude, the project maintains and even improves its modular decomposition, as is clearly shown in the Figures \ref{fig:jena:01} - \ref{fig:jena:03}. From a software engineering perspective, the structure of \textsc{Jena} shown in Figure \ref{fig:jena:03} is favorable, since it allows for an easy decomposition, maintenance and replacement of individual packages. One of the possible reasons for the discrepancy between \textsc{Jena} and \textsc{Azureus} is that the first is a framework aimed at an audience of developers. Thus, its structure must be well organized to facilitate its adoption, while the second is an end-user application and therefore the focus is on functionality rather than structural quality.

We are currently working on the extension of our approach in a way in which we hope to uncover the full potential of the $Q$-metric and its correlation with other software development processes, by modeling this dynamics as a simple network growth process with an underlying modular decomposition. This is the subject of ongoing research \cite{zanettimodelmodularity}. Along the way we aim at improving our research methodology with more insights based on the network science framework as well as aligning it with existing results from the software engineering community. Prior to concluding this article and giving details on future research, in the next section we comment on related work.
      
\section{Related Work}\label{rel} 
 
One of the eye catching features of the time evolution of the $Q$-metric, as presented in Figure \ref{f:allqtimeav}, is the large fluctuation of $Q$ at early stages of the project development. This is in accordance with results reported in \cite{sustainablegrowthtessone2011}. There, it was shown that young open source software projects display an accelerated growth rate while mature projects stabilize their dynamics and can grow further in a sustainable regime. 

Another possible, complementary, reason for fluctuations are refactoring events, where software is usually rewritten or restructured in order to improve multiple features such as functionality, flexibility, reusability or structural quality. Such events could lead to the sudden jumps observed in Figure \ref{f:allqtimeav} along the time evolution of a software project. In \cite{findingrefactoring2000}, refactoring metrics are proposed which take into account the dynamics of changing code. This line of research is well aligned with our purposes and can be easily adapted and augmented by our network perspective on software development processes. 

For an early attempt of the application of network science to the analysis of software engineering processes we recommend \cite{presoftwarenetworks2003}, which also contains a short review of classical approaches used in the software engineering literature. Finally, a recent article published in the \textsc{PNAS} journal used a similar network approach, though with a different metric, to study modularity of code and its relation to module survival, drawing a parallel to ecological systems and making use of a predator-prey model variation \cite{softwareevolutionpnas2011}.
   
\section{Conclusion and Outlook}\label{conc} 
 
The results presented in section \ref{res} indicate that the $Q$-metric known from the analysis of cluster structures in network science is a promising and reasonable approach to quantify the coherence between the package decomposition of large software projects and their dependency structures. As such, it constitutes a macroscopic measure that allows us to monitor and evaluate software engineering processes and reason about the sustainability of software architectures. In particular, it provides a simple mapping from local development activities to their respective impact on the mesoscopic and macroscopic structures of software systems. 
Although the current metric offers interesting insights, a known issue is that it is being influenced by intra-module dependencies. However it would be more thoughtful to place more weight on the impact of inter-module dependencies because these are the most relevant dependencies in a software modular structure. Last but not least, \textsc{Java} packages which were used as proxy for modularity in \textsc{Java} source code have a hierarchical structure. Therefore, dependencies between packages $a.b.c.d$ and $a.b.c.e$ are of less concern than between packages $a.b.c.d$ and $x.y.z$. 
            
While all these issues are the subject to future investigations, our study already foreshadows a number of interesting research questions: How does the evolution of $Q$ impact the sustainability of distributed software engineering efforts? Can the incorporation of such macroscopic measures into software development tools improve the design and maintenance of software architectures? How is the dynamics of $Q$ over the lifetime of software projects correlated with software development acts like refactoring or bug fixing? How is it correlated with social aspects, coordination acts or communication processes taking place between developers? Intuitively, one would assume that a reasonable modular decomposition of complex software systems facilitates distributed development processes and mitigates change propagation between interdependent modules. An interesting future work is thus to augment the results in this paper with data on coordination and communication acts in the respective projects. In this line of arguments, a further interesting question is whether the pronouncedness of modular structures in the dependency network allows us to infer statements about the hierarchical organization of development teams.

While the exploration of these questions in this study has been necessarily incomplete, we argue that the associated line of research is a good demonstration for the potential impact of complex systems science on the engineering of complex software systems.
  
\section*{Acknowledgment}   
   
We acknowledge the financial support provided by the Swiss National Science Foundation through grant CR12I1\_125298 and also Ingo Scholtes, Claudio Juan Tessone and our three reviewers for valuable comments.
 
\bibliography{ref}

\end{document}